\documentclass[reprint,prd,twocolumn,superscriptaddress]{revtex4-1}
\pdfoutput=1
\usepackage{graphics,times,mathptmx}
\newcommand{\F}[4]{\,_{#1}F_{#2}\left(\left.\begin{array}{c}#3\end{array}\right|#4\right)}
\newcommand{\Li}[1]{\mathop{\mathrm{Li}}\nolimits_{#1}}

\begin{document}
\title{HQET vertex diagram: $\varepsilon$ expansion}
\author{Andrey G.~Grozin}
\email{A.G.Grozin@inp.nsk.su}
\affiliation{Budker Institute of Nuclear Physics, Lavrentyev St.~11, Novosibirsk 630090, Russia}
\affiliation{Novosibirsk State University, Pirogov St.~1, Novosibirsk 630090, Russia}
\date{}
\begin{abstract}
Differential equations for the one-loop HQET vertex diagram
with arbitrary self-energy insertions and arbitrary residual energies
are reduced to the $\varepsilon$ form
and used to obtain the $\varepsilon$ expansion in terms of Goncharov polylogarithms.
\end{abstract}
\maketitle

\begin{figure}
\includegraphics{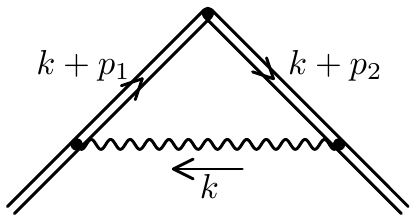}
\caption{The one-loop HQET vertex diagram}
\label{Diagram}
\end{figure}

We consider the one-loop vertex diagram (Fig.~\ref{Diagram})
with arbitrary degrees of all 3 denominators:
\begin{eqnarray}
&&I_{n_1,n_2,n_3}(\vartheta;\omega_1,\omega_2) = \frac{1}{i\pi^{d/2}}
\int \frac{d^d k}{D_1^{n_1} D_2^{n_2} D_3^{n_3}}\,,
\nonumber\\
&&D_1 = - 2 (k + p_1) \cdot v_1\,,\quad
D_2 = - 2 (k + p_2) \cdot v_2\,,
\nonumber\\
&&D_3 = - k^2\,,
\label{Definition}
\end{eqnarray}
where $\omega_{1,2} = p_{1,2} \cdot v_{1,2}$, $\cosh\vartheta=v_1 \cdot v_2$.
It has obvious properties
\begin{eqnarray}
&&I_{n_1,n_2,n_3}(\vartheta;\omega_1,\omega_2) =
I_{n_1,n_2,n_3}(-\vartheta;\omega_1,\omega_2)\,,
\label{Sym0}\\
&&I_{n_1,n_2,n_3}(\vartheta;\omega_1,\omega_2) =
I_{n_2,n_1,n_3}(\vartheta;\omega_2,\omega_1)\,,
\label{Sym1}\\
&&I_{n_1,0,n_3}(\vartheta;\omega_1,\omega_2) =
I_{n_1,n_3} (-2\omega_1)^{d-n_1-2n_3}\,,
\label{n20}
\end{eqnarray}
where
\begin{equation}
I_{n_1,n_2} = \frac{\Gamma(n_1+2n_2-d) \Gamma(d/2-n_2)}{\Gamma(n_1) \Gamma(n_2)}
\label{selfen}
\end{equation}
is the one-loop HQET self-energy diagram.

Results exact in $\varepsilon$ are known for $\omega_1 = \omega_2$~\cite{Grozin:2011rs}
\begin{eqnarray}
&&I_{n_1,n_2,n_3}(\vartheta;\omega,\omega) =
I_{n_1+n_2,n_3} (-2\omega)^{d-n_1-n_2-2n_3}
\nonumber\\
&&{}\times
\F{3}{2}{n_1,n_2,\frac{d}{2}-n_3\\\frac{n_1+n_2}{2},\frac{n_1+n_2+1}{2}}{\frac{1 - \cosh\vartheta}{2}}
\label{F32}
\end{eqnarray}
and $\vartheta = 0$~\cite{Bagan:1992ty}
\begin{eqnarray}
&&I_{n_1,n_2,n_3}(0;\omega_1,\omega_2)
= I_{n_1+n_2,n_3} (-2\omega_2)^{d-n_1-n_2-2n_3}
\nonumber\\
&&{}\times\F{2}{1}{n_1,n_1+n_2+2n_3-d\\n_1+n_2}{1-y}\,.
\label{BBG}
\end{eqnarray}
(the symmetry~(\ref{Sym1}) follows from a hypergeometric identity).
Here and below we use $d = 4 - 2 \varepsilon$,
\begin{equation}
x = e^\vartheta\,,\quad
y = \frac{\omega_1}{\omega_2}\,.
\label{xy}
\end{equation}

We consider the one-loop vertex (Fig.~\ref{Diagram})
with any numbers of self-energy insertions into each of 3 lines,
provided that all lines in these insertions are massless.
If the full number of loops in all self-energy insertions into the line $i$ is $l_i$,
then $n_{1,2} = m_{1,2} + 2 l_{1,2} \varepsilon$, $n_3 = m_3 + l_3 \varepsilon$,
where all $m_i$ are integer.
All integrals with a given set $l_i$ can be reduced~\cite{Grozin:2011rs}, using IBP,
to 3 master integrals with $m_i = (0,1,1)$, $(1,0,1)$ and $(1,1,1)$.
We choose the column of the basis integrals $(f_1,f_2,f_3)^T$, where
\begin{eqnarray}
&&I_{2 l_1 \varepsilon, 1 + 2 l_2 \varepsilon, 1 + l_3 \varepsilon}(\vartheta;\omega_1,\omega_2)
= I_{1+2(l_1+l_2)\varepsilon,1+l_3\varepsilon}
\nonumber\\
&&\quad{} \times (-2\omega_1)^{-l\varepsilon} (-2\omega_2)^{1-l\varepsilon} f_1(x,y)\,,
\nonumber\\
&&I_{1 + 2 l_1 \varepsilon, 2 l_2 \varepsilon, 1 + l_3 \varepsilon}(\vartheta;\omega_1,\omega_2)
= I_{1+2(l_1+l_2)\varepsilon,1+l_3\varepsilon}
\nonumber\\
&&\quad{} \times (-2\omega_1)^{1-l\varepsilon} (-2\omega_2)^{-l\varepsilon} f_2(x,y)\,,
\nonumber\\
&&I_{1 + 2 l_1 \varepsilon, 1 + 2 l_2 \varepsilon, 1 + l_3 \varepsilon}(\vartheta;\omega_1,\omega_2)
= I_{2+2(l_1+l_2)\varepsilon,1+l_3\varepsilon}
\nonumber\\
&&\quad{} \times (-2\omega_1)^{-l\varepsilon} (-2\omega_2)^{-l\varepsilon} f_3(x,y)\,,
\label{f}
\end{eqnarray}
where $l=l_1+l_2+l_0$ is the total number of loops, $l_0=l_3+1$.
They have symmetry properties
\begin{eqnarray}
&&f(x^{-1},y) = f(x,y)\,,
\label{Sym}\\
&&f(x,y^{-1}) = S_y \left[f(x,y)\right]_{l_1\leftrightarrow l_2}\,,\quad
S_y = \left(\begin{array}{rrr}0&1&0\\1&0&0\\0&0&1\end{array}\right)\,.
\nonumber
\end{eqnarray}
The initial condition is $f(1,1) = (1,1,1)^T$.
If $l_1=0$, $f_1$ is trivial~(\ref{n20});
if $l_2=0$, $f_2$ is trivial;
if $l_1=l_2=0$, there is only one non-trivial master integral $f_3$.
If $l_1=l_2$, $f_2(x,y)=f_1(x,y^{-1})$~(\ref{Sym}),
and there are only 2 unknown functions $f_1$ and $f_3$.

We shall use the method of differential equations~\cite{Kotikov:1990kg}.
Using
\begin{eqnarray}
&&\sinh\vartheta \frac{\partial}{\partial\vartheta} I_{n_1,n_2,n_3}
\nonumber\\
&&{} = n_1 \left[\mathbf{1}^+ \mathbf{2}^- - 2 (\omega_1 \cosh\vartheta - \omega_2) \mathbf{1}^+ - \cosh\vartheta\right] I_{n_1,n_2,n_3}
\nonumber\\
&&{} = n_2 \left[\mathbf{2}^+ \mathbf{1}^- - 2 (\omega_2 \cosh\vartheta - \omega_1) \mathbf{2}^+ - \cosh\vartheta\right] I_{n_1,n_2,n_3}\,,
\nonumber\\
&&\frac{\partial}{\partial\omega_1} I_{n_1,n_2,n_3} =
2 n_1 \mathbf{1}^+ I_{n_1,n_2,n_3}\,,
\nonumber\\
&&\frac{\partial}{\partial\omega_2} I_{n_1,n_2,n_3} =
2 n_2 \mathbf{2}^+ I_{n_1,n_2,n_3}
\label{de0}
\end{eqnarray}
and the IBP reduction, we can derive the differential equations
\begin{equation}
\partial_x f = M_x f\,,\quad
\partial_y f = M_y f\,,
\label{de1}
\end{equation}
where the matrices $M_{x,y}$ (depending on $x$, $y$ and $\varepsilon$) satisfy
\begin{equation}
\partial_x M_y - \partial_y M_x - [M_x,M_y] = 0
\label{cons}
\end{equation}
because $\partial_x \partial_y f = \partial_y \partial_x f$.
The symmetries~(\ref{Sym}) lead to
\begin{eqnarray}
&&M_x(x^{-1},y) + x^2 M_x(x,y) = 0\,,\quad
M_y(x^{-1},y) = M_y(x,y)\,;
\nonumber\\
&&M_x(x,y^{-1}) = S_y \left[M_x\right]_{l_1\leftrightarrow l_2} S_y\,,
\nonumber\\
&&M_y(x,y^{-1}) + y^2 S_y \left[M_y\right]_{l_1\leftrightarrow l_2} S_y = 0\,.
\label{SymM}
\end{eqnarray}

The differential equations~(\ref{de1}) can be reduced to the canonical form~\cite{Henn:2013pwa}
by a linear transformation $f = T F$ (the matrix $T$ depends on $x$, $y$, $\varepsilon$),
\begin{equation}
d F = \varepsilon\,d M\,F\,,\quad
M(x,y) = \sum_i M_i \log p_i(x,y)\,,
\label{de2}
\end{equation}
where $p_i(x,y)$ are polynomials in $x$ and $y$, and $M_i$ are constant matrices.
We use the Mathematica package Libra~\cite{Libra}
which implements the algorithm of~\cite{Lee:2014ioa},
and obtain
\begin{eqnarray}
&&T = \left(\begin{array}{ccc}
1 & 0 & l_1 \frac{1 + x^2 - 2 x y}{1 - x^2} \\
0 & 1 & l_2 \frac{1 + x^2 - 2 x y^{-1}}{1 - x^2} \\
0 & 0 & - \frac{1 + 2 (l_1+l_2) \varepsilon}{\varepsilon} \frac{x}{1 - x^2}
\end{array}\right)\,,
\nonumber\\
&&T^{-1} = \left(\begin{array}{ccc}
1 & 0 & \frac{l_1 \varepsilon}{1+2(l_1+l_2)\varepsilon} \frac{1+x^2-2xy}{x} \\
0 & 1 & \frac{l_2 \varepsilon}{1+2(l_1+l_2)\varepsilon} \frac{1+x^2-2xy^{-1}}{x} \\
0 & 0 & - \frac{\varepsilon}{1 + 2 (l_1+l_2) \varepsilon} \frac{1 - x^2}{x}
\end{array}\right)\,.
\label{T}
\end{eqnarray}
The symmetry properties of the canonical master integrals are
\begin{eqnarray}
&&F(x^{-1},y) = S_x F(x,y)\,,\quad
S_x = \left(\begin{array}{rrr}1&0&0\\0&1&0\\0&0&-1\end{array}\right)\,,
\nonumber\\
&&F(x,y^{-1}) = S_y \left[F(x,y)\right]_{l_1\leftrightarrow l_2}\,.
\label{Sym2}
\end{eqnarray}
The initial conditions for the differential equations~(\ref{de2}) are
\begin{equation}
F(1,1) = T^{-1}(1,1) f(1,1) = (1,1,0)^T\,.
\label{Init}
\end{equation}
The matrix $M(x,y)$ is
\begin{eqnarray}
&&M = M_1 \log x + M_2 \left[\log(1+x) + \log(1-x)\right] + M_3 \log y
\nonumber\\
&&\quad{} + M_4 \log(x-y) + M_5 \log(1-xy)\,,
\label{dlog}\\
&&M_1 = \left(\begin{array}{ccc}
l_1 & -l_1 & l_1 (l_1-l_2+l_0) \\
-l_2 & l_2 & l_2 (-l_1+l_2+l_0) \\
1 & 1 & l_1+l_2-l_0
\end{array}\right)\,,
\nonumber\\
&&M_2 = \left(\begin{array}{ccc}
0 & 0 & 0 \\
0 & 0 & 0 \\
0 & 0 & 2 l_0
\end{array}\right)\,,\quad
M_3 = \left(\begin{array}{ccc}
l & 0 & 0 \\
- 2 l_2 & -l_1+l_2-l_0 & 0 \\
0 & 0 & l
\end{array}\right)\,,
\nonumber\\
&&M_4 = \left(\begin{array}{ccc}
-l_1 & l_1 & l_1 l \\
l_2 & -l_2 & - l_2 l \\
1 & -1 & -l
\end{array}\right)\,,\quad
M_5 = \left(\begin{array}{ccc}
-l_1 & l_1 & - l_1 l \\
l_2 & -l_2 & l_2 l \\
-1 & 1 & -l
\end{array}\right)
\nonumber
\end{eqnarray}
(only derivatives of $M$ matter,
and hence we may freely substitute $\log(y-x) \to \log(x-y)$, etc.).
This matrix has symmetry properties
\begin{eqnarray}
&&M(x^{-1},y) = S_x M(x,y) S_x\,,
\nonumber\\
&&M(x,y^{-1}) = S_y \left[M(x,y)\right]_{l_1 \leftrightarrow l_2} S_y
\label{SymM2}
\end{eqnarray}
(again, up to inessential additive constants).

If $l_1=0$ then $F_1(x,y) = y^{l\varepsilon}$.
The first equation decouples, and this trivial function satisfies this equation.
The two non-trivial master integrals $F_{2,3}$ are determined by coupled equations.
The case $l_2=0$ is similar.
If $l_1=l_2=0$ then $F_{1,2}(x,y) = y^{\pm l\varepsilon}$;
the only non-trivial master integral $F_3$ is determined by the third equation.

First we consider the single-scale case $y=1$.
The differential equations for $x<1$ are
\begin{eqnarray}
&&\frac{d F(x,1)}{d x} = \varepsilon
\left[\frac{M_1}{x} + \frac{M_2}{x+1} + \frac{M_2+M_4+M_5}{x-1}\right]
F(x,1)\,,
\nonumber\\
&&M_2 + M_4 + M_5 = 2 \left(\begin{array}{ccc}
-l_1 & l_1 & 0 \\
l_2 & -l_2 & 0 \\
0 & 0 & -l_1-l_2
\end{array}\right)\,.
\label{dex1}
\end{eqnarray}
For $x>1$ we have $F(x,1) = S_x F(x^{-1},1)$;
these functions satisfy the equations
\begin{eqnarray*}
&&\frac{d F(x,1)}{d x^{-1}} = \varepsilon
\biggl[ - \frac{M_1 + 2 M_2 + M_4 + M_5}{x^{-1}}
\nonumber\\
&&{} + \frac{M_2}{x^{-1}+1} + \frac{M_2+M_4+M_5}{x^{-1}-1}\biggr]
F(x,1)
\end{eqnarray*}
because $- S_x (M_1 + 2 M_2 + M_4 + M_5) S_x = M_1$,
$S_x M_2 S_x = M_2$, $S_x (M_2 + M_4 + M_5) S_x = M_2 + M_4 + M_5$
(this follows from~(\ref{SymM2})).

The solution of the differential equations~(\ref{dex1}) with the initial conditions~(\ref{Init})
as a series in $\varepsilon$ can be obtained using Libra.
The coefficients are uniform-weight combinations of harmonic polylogarithms~\cite{Remiddi:1999ew}
(we use HPL~\cite{Maitre:2005uu,Maitre:2007kp} to reduce them to a minimal set):
\begin{widetext}
\begin{eqnarray}
&&F_1(x,1) = 1
+ l_1 (l_1-l_2+l_0) H_{0}^2(x) \varepsilon^2
\nonumber\\
&&{} + 2 l_1 \biggl\{
(l_1-l_2+l_0) \biggl[
- 4 l_0 H_{0,0,-1}(x) + 2 l_0 H_{0}(x) H_{0,-1}(x) + (2 l - l_0) \frac{\pi^2}{6} H_{0}(x)
\biggr]
- 2 \bigl(l_1^2 - l_2^2 + (l_1 + 3 l_2) l_0\bigr) H_{0,0,1}(x)
\nonumber\\
&&\quad{} + 4 l_2 l_0 H_{0}(x) H_{0,1}(x)
+ (l_1-l_2) l H_{0}^2(x) H_{1}(x)
+ \frac{1}{6} \bigl(2 (l_1^2-l_2^2) + (l_1+l_2) l_0 - l_0^2\bigr) H_{0}^3(x)
+ \bigl(2 (l_1^2-l_2^2) + (5 l_1 + 3 l_2) l_0 + 3 l_0^2\bigr) \zeta_3
\biggr\} \varepsilon^3
\nonumber\\
&&{} + 2 l_1 \biggl\{
(l_1-l_2) l \biggl[
2 l_0 \bigl(4 H_{0,0,1,-1}(x) + 4 H_{0,0,-1,1}(x) + 2 H_{0,1,0,-1}(x) - 3 H_{0,0,0,-1}(x)
- 4 H_{1}(x) H_{0,0,-1}(x) + 2 H_{0}(x) H_{1}(x) H_{0,-1}(x)
\bigr)
\nonumber\\
&&\qquad{} - (l_1+l_2) \bigl(2 H_{0,1,0,1}(x) + 4 H_{1}(x) H_{0,0,1}(x) - H_{0}^2(x) H_{1}^2(x)\bigr)
- \bigl(2 (l_1+l_2) - l_0\bigr)
\biggl(2 H_{0,0,0,1}(x) - \frac{1}{3} H_{0}^3(x) H_{1}(x)\biggr)
\nonumber\\
&&\qquad{} + (2 l - l_0) \frac{\pi^2}{3} H_{0}(x) H_{1}(x)
+ 2 (2 l + l_0) \zeta_3 H_{1}(x)
\biggr]
\nonumber\\
&&\quad{} + (l_1-l_2+l_0) \biggl[
4 l_0^2 H_{0}(x) H_{0,-1,-1}(x) - 2 l_0^2 H_{0,-1}^2(x) - 4 (l_1+l_2) l_0 H_{0,1}(x) H_{0,-1}(x)
+ (l_1+l_2-l_0) l_0 H_{0}^2(x) H_{0,-1}(x)
\nonumber\\
&&\qquad{} + (2 l - l_0) l_0 \frac{\pi^2}{3} H_{0,-1}(x)
\biggr]
+ (l_1+l_2) \bigl[
8 l_2 l_0 H_{0}(x) H_{0,1,1}(x) + \bigl(l_1^2 - l_2^2 + (l_1 - 5 l_2) l_0\bigr) H_{0,1}^2(x)
\bigr]
\nonumber\\
&&\quad{} + 8 l_2 l_0^2 H_{0}(x) \bigl(H_{0,1,-1}(x) + H_{0,-1,1}(x)\bigr)
- 2 l_2 (l_1+l_2-l_0) l_0 H_{0}(x) \bigl(2 H_{0,0,1}(x) - H_{0}(x) H_{0,1}(x)\bigr)
\nonumber\\
&&\quad{} + 2 (l_1 - 3 l_2 + l_0) l_0^2 H_{0}(x) H_{0,0,-1}(x)
+ \frac{1}{24} \bigl(4 (l_1^3-l_2^3) + 2 (l_1^2+l_2^2) l_0 - (l_1+l_2) l_0^2 + l_0^3\bigr) H_{0}^4(x)
\nonumber\\
&&\quad{} + (2 l - l_0) \frac{\pi^2}{12} \biggl[
8 l_2 l_0 H_{0,1}(x) + \bigl(2 (l_1^2-l_2^2) + (l_1+l_2) l_0 - l_0^2\bigr) H_{0}^2(x)
+ \bigl(6 (l_1^2-l_2^2) + (l_1 - 21 l_2) l_0 - 5 l_0^2\bigr) \frac{\pi^2}{30}
\biggr]
\nonumber\\
&&\quad{} - 2 \bigl(2 l_2 l + l_1 l_0 + l_0^2\bigr) l_0 \zeta_3 H_{0}(x)
\biggr\} \varepsilon^4 + \mathcal{O}(\varepsilon^5)\,,
\nonumber\\
&&F_2(x,1) = \left[F_1(x,1)\right]_{l_1\leftrightarrow l_2}\,,
\nonumber\\
&&F_3(x,1) = \left[F_3(x,1)\right]_{l_1\leftrightarrow l_2} = 2 H_{0}(x) \varepsilon
\nonumber\\
&&{} + \biggl[
- 4 (l_1+l_2) \bigl(H_{0,1}(x) - H_{0}(x) H_{1}(x)\bigr)
- 4 l_0 \bigl(H_{0,-1}(x) - H_{0}(x) H_{-1}(x)\bigr)
+ (l_1+l_2-l_0) H_{0}^2(x)
+ (2 l - l_0) \frac{\pi^2}{3}
\biggr] \varepsilon^2
\nonumber\\
&&{} + \biggl\{
4 (l_1+l_2)^2 \bigl(2 H_{0,1,1}(x) - 2 H_{1}(x) H_{0,1}(x) + H_{0}(x) H_{1}^2(x)\bigr)
\nonumber\\
&&\quad{} + 2 (l_1+l_2) \biggl[
4 l_0 \bigl(H_{0,1,-1}(x) + H_{0,-1,1}(x) - H_{-1}(x) H_{0,1}(x) - H_{1}(x) H_{0,-1}(x)
+ H_{0}(x) H_{1}(x) H_{-1}(x)\bigr)
\nonumber\\
&&\qquad{} - (l_1+l_2-l_0) \bigl(2 H_{0,0,1}(x) - H_{0}^2(x) H_{1}(x)\bigr)
+ (2 l - l_0) \frac{\pi^2}{3} H_{1}(x)
\biggr]
\nonumber\\
&&\quad{} + 4 l_0^2 \bigl(2 H_{0,-1,-1}(x) - 2 H_{-1}(x) H_{0,-1}(x) + H_{0}(x) H_{-1}^2(x)\bigr)
- (l_1+l_2-l_0) \biggl[
4 l_0 H_{0,0,-1}(x) - 2 l_0 H_{0}^2(x) H_{-1}(x) - (2 l - l_0) \frac{\pi^2}{3} H_{0}(x)
\biggr]
\nonumber\\
&&\quad{} + \frac{1}{3} \bigl(2 (l_1^2+l_2^2) - (l_1+l_2) l_0 + l_0^2\bigr) H_{0}^3(x)
+ 2 (2 l - l_0) l_0 \frac{\pi^2}{3} H_{-1}(x)
- 2 \bigl(2 (l_1+l_2)^2 + 3 (l_1+l_2) l_0 + 2 l_0^2\bigr) \zeta_3
\biggr\} \varepsilon^3
\nonumber\\
&&{} + \biggl\{
8 (l_1+l_2)^3 \biggl(
- 2 H_{0,1,1,1}(x) + 2 H_{1}(x) H_{0,1,1}(x) - H_{1}^2(x) H_{0,1}(x) + \frac{1}{3} H_{0}(x) H_{1}^3(x)
\biggr)
\nonumber\\
&&\quad{} - 2 (l_1+l_2)^2 \biggl[
(l_1+l_2-l_0) \bigl(2 H_{0,1,0,1}(x) + 4 H_{1}(x) H_{0,0,1}(x)
- H_{0,1}^2(x) - H_{0}^2(x) H_{1}^2(x)\bigr)
\nonumber\\
&&\qquad{} + 4 l_0 \bigl(
2 H_{0,1,1,-1}(x) + 2 H_{0,1,-1,1}(x) + 2 H_{0,-1,1,1}(x)
- 2 H_{-1}(x) H_{0,1,1}(x) - 2 H_{1}(x) H_{0,1,-1}(x) - 2 H_{1}(x) H_{0,-1,1}(x)
\nonumber\\
&&\qquad\quad{} + 2 H_{1}(x) H_{-1}(x) H_{0,1}(x) + H_{1}^2(x) H_{0,-1}(x) - H_{0}(x) H_{-1}(x) H_{1}^2(x)
\bigr)
- (2 l - l_0) \frac{\pi^2}{3} H_{1}^2(x)
\biggr]
\nonumber\\
&&\quad{} - 8 l_0^3 \biggl(
2 H_{0,-1,-1,-1}(x) - 2 H_{-1}(x) H_{0,-1,-1}(x) + H_{-1}^2(x) H_{0,-1}(x)
- \frac{1}{3} H_{0}(x) H_{-1}^3(x)
\biggr)
\nonumber\\
&&\quad{} + 2 (l_1+l_2-l_0) \biggl[
(l_1+l_2) \biggl(
2 l_0 \bigl(2 H_{0,0,1,-1}(x) + 2 H_{0,0,-1,1}(x) - 2 H_{-1}(x) H_{0,0,1}(x) - 2 H_{1}(x) H_{0,0,-1}(x)
+ H_{0}^2(x) H_{1}(x) H_{-1}(x)
\bigr)
\nonumber\\
&&\qquad{} + (2 l - l_0) \frac{\pi^2}{3} H_{0}(x) H_{1}(x)
\biggr)
+ l_0^2 \bigl(4 H_{0,0,-1,-1}(x) - 4 H_{-1}(x) H_{0,0,-1}(x) + H_{0}^2(x) H_{-1}^2(x)\bigr)
+ (2 l - l_0) l_0 \frac{\pi^2}{3} H_{0}(x) H_{-1}(x)
\biggr]
\nonumber\\
&&\quad{} - 2 (l_1+l_2) \biggl[
4 l_0^2 \bigl(2 H_{0,1,-1,-1}(x) + 2 H_{0,-1,1,-1}(x) + 2 H_{0,-1,-1,1}(x)
- 2 H_{-1}(x) H_{0,1,-1}(x) - 2 H_{-1}(x) H_{0,-1,1}(x)
\nonumber\\
&&\qquad\quad{} - 2 H_{1}(x) H_{0,-1,-1}(x)
+ H_{-1}^2(x) H_{0,1}(x) + 2 H_{1}(x) H_{-1}(x) H_{0,-1}(x)
- H_{0}(x) H_{1}(x) H_{-1}^2(x)
\bigr)
- 2 (2 l - l_0) l_0 \frac{\pi^2}{3} H_{1}(x) H_{-1}(x)
\biggr]
\nonumber\\
&&\quad{} - 4 \bigl(2 (l_1+l_2) (l_1^2+l_2^2) - (l_1^2 + l_2^2 - 10 l_1 l_2) l_0 + (l_1+l_2) l_0^2\bigr)
H_{0,0,0,1}(x)
- 4 \bigl(5 (l_1^2+l_2^2) - 6 l_1 l_2 + 2 (l_1+l_2) l_0 + l_0^2\bigr) l_0 H_{0,0,0,-1}(x)
\nonumber\\
&&\quad{}+ 8 l_1 l_2 l_0 H_{0}(x) \bigl(4 H_{0,0,1}(x) - H_{0}(x) H_{0,1}(x)\bigr)
+ 2 \bigl((l_1-l_2)^2 + (l_1+l_2) l_0\bigr) l_0 H_{0}(x)
\bigl(4 H_{0,0,-1}(x) - H_{0}(x) H_{0,-1}(x)\bigr)
\nonumber\\
&&\quad{} + \frac{1}{3} \bigl(2 (l_1^2+l_2^2) - (l_1+l_2) l_0 + l_0^2\bigr) H_{0}^2(x) \biggl[
2 (l_1+l_2) H_{0}(x) H_{1}(x) + 2 l_0 H_{0}(x) H_{-1}(x) + (2 l - l_0) \frac{\pi^2}{2}
\biggr]
\nonumber\\
&&\quad{} + \frac{1}{12} \bigl(4 (l_1^3+l_2^3) - 2 (l_1^2+l_2^2) l_0 + (l_1+l_2) l_0^2 - l_0^3\bigr) H_{0}^4(x)
\nonumber\\
&&\quad{} + (2 l - l_0) \frac{\pi^2}{3} \biggl[
2 l_0^2 H_{-1}^2(x) + \bigl(22 (l_1^2+l_2^2) + 28 l_1 l_2 + 13 (l_1+l_2) l_0 + 9 l_0^2\bigr)
\frac{\pi^2}{60}
\biggr]
\nonumber\\
&&\quad{} - 4 \bigl(2 (l_1+l_2)^2 + 3 (l_1+l_2) l_0 + 2 l_0^2\bigr) \zeta_3
\bigl[(l_1+l_2) H_{1}(x) + l_0 H_{-1}(x)\bigr]
\nonumber\\
&&\quad{} - 4 \bigl(4 l_1 l_2 (l_1+l_2) - 2 (l_1^2 + l_2^2 + l_1 l_2) l_0 - 2 (l_1+l_2) l_0^2 - l_0^3\bigr)
\zeta_3 H_{0}(x)
\biggr\} \varepsilon^4 + \mathcal{O}(\varepsilon^5)\,.
\label{Fx}
\end{eqnarray}
\end{widetext}
This expansion can be straightforwardly extended to any order in $\varepsilon$.
We have also expanded the exact hypergeometric representations of $F_{1,3}(x,1)$
which follow from~(\ref{F32}) up to $\varepsilon^3$ using HypExp~\cite{Huber:2005yg,Huber:2007dx}.
The results can be expressed via ordinary polylogarithms up to $\Li3$,
and agree with~(\ref{Fx}).
They also agree with the expansions up to $\varepsilon^3$ obtained in~\cite{Grozin:2011rs}
(also using~(\ref{F32}) and HypExp).
When $l_1=l_2=0$, the only non-trivial master integral $F_3$
is expressed~(\ref{F32}) via the ${}_2F_1$ function
whose $\varepsilon$ expansion is known to all orders~\cite{Davydychev:2000na}.
The expansion in euclidean case is given there
(there is a typo in the journal version corrected in the version 4 in arXiv);
the Minkowski case is given by the formula~(41) in~\cite{Grozin:2017aty}.
Our result~(\ref{Fx}) at $l_1=l_2=0$ agrees with the formula~(B.10) in~\cite{Grozin:2017aty}
(it contains 3 further expansion terms).

Any finite number of terms in the expansion of $F(x,1)$ in $\bar{x}=1-x$
can be straightforwardly obtained from~(\ref{F32}):
\begin{eqnarray}
&&F_1(x,1) = 1
+ \frac{\varepsilon^2 l_1 \bar{x}^2}{(1+(l_1+l_2)\varepsilon) (1+2(l_1+l_2)\varepsilon)}
\nonumber\\
&&\left\{(l_1 - l_2 + l_0 + 2 l_2 l_0 \varepsilon)
(1 + \bar{x}) + \mathcal{O}(\bar{x}^2)\right\}\,,
\nonumber\\
&&F_2(x,1) = \left[F_1(x,1)\right]_{l_1\leftrightarrow l_2}\,,
\nonumber\\
&&F_3(x,1) = - \frac{\varepsilon \bar{x}}{1+2(l_1+l_2)\varepsilon}
\left[2 + \bar{x} + \mathcal{O}(\bar{x}^2)\right]
\label{Ser1x}
\end{eqnarray}
(we have obtained them up to $\bar{x}^{20}$).
The coefficients are exact functions of $\varepsilon$.
This expansion satisfies the differential equation~(\ref{dex1})
with the initial condition~(\ref{Init}).
Expanding each coefficient of~(\ref{Ser1x}) in $\varepsilon$,
and each coefficient of~(\ref{Fx}) in $\bar{x}$,
we obtain two identical double expansions up to $\varepsilon^4$ and $\bar{x}^{20}$;
this is a strong check of our result~(\ref{Fx}).

Next we consider the straight-line case $x=1$.
From the form of the matrix $T^{-1}$~(\ref{T}) at $x=1$ we see that $F_3(1,y) = 0$.
The differential equations for $y<1$ are
\begin{eqnarray}
&&\frac{d F(1,y)}{d y} = \varepsilon
\left[\frac{M_3}{y} + \frac{M_4+M_5}{y-1}\right]
F(1,y)\,,
\nonumber\\
&&M_4 + M_5 = 2 \left(\begin{array}{ccc}
-l_1 & l_1 & 0 \\
l_2 & -l_2 & 0 \\
0 & 0 & -l
\end{array}\right)
\label{dey1}
\end{eqnarray}
(they are, of course, consistent with $F_3=0$).
For $y>1$ we have $F(1,y) = S_y \left[F(1,y^{-1})\right]_{l_1\leftrightarrow l_2}$;
these functions satisfy the equations
\[
\frac{d F(1,y)}{d y^{-1}} = \varepsilon
\biggl[ - \frac{M_3 + M_4 + M_5}{y^{-1}} + \frac{M_4+M_5}{y^{-1}-1}\biggr]
F(1,y)
\]
because $- S_y \left[M_3 + M_4 + M_5\right]_{l_1\leftrightarrow l_2} S_y = M_3$,
$S_y \left[M_4 + M_5\right]_{l_1\leftrightarrow l_2} S_y = M_4 + M_5$
(this follows from~(\ref{SymM2})).

Solving the differential equations~(\ref{dey1}) with the initial conditions~(\ref{Init}) we obtain
\begin{widetext}
\begin{eqnarray}
&&y^{-l\varepsilon} F_1(1,y) = 1
- 4 l_1 l \biggl(H_{0,1}(y) - H_{0}(y) H_{1}(y) - \frac{\pi^2}{6}\biggr) \varepsilon^2
\nonumber\\
&&{} + 4 l_1 l \biggl[
(l_1+l_2) \biggl(2 H_{0,1,1}(y) - 2 H_{1}(y) H_{0,1}(y) + H_{0}(y) H_{1}^2(y)
+ \frac{\pi^2}{3} H_{1}(y)\biggr)
\nonumber\\
&&\quad{} - (l_1+l_0) \bigl(2 H_{0,0,1}(y) - 2 H_{0}(y) H_{0,1}(y) + H_{0}^2(y) H_{1}(y)\bigr)
- 2 (l_2-l_0) \zeta_3
\biggr] \varepsilon^3
\nonumber\\
&&{} - 4 l_1 l \biggl[
2 (l_1+l_2)^2 \biggl(2 H_{0,1,1,1}(y) - 2 H_{1}(y) H_{0,1,1}(y) + H_{1}^2(y) H_{0,1}(y)
- \frac{1}{3} H_{0}(y) H_{1}^3(y) - \frac{\pi^2}{6} H_{1}^2(y)\biggr)
\nonumber\\
&&\quad{} + 2 (l_1+l_0)^2 \biggl(2 H_{0,0,0,1}(y) - 2 H_{0}(y) H_{0,0,1}(y) + H_{0}^2(y) H_{0,1}(y)
- \frac{1}{3} H_{0}^3(y) H_{1}(y)\biggr)
\nonumber\\
&&\quad{} + 2 (l_1 l - l_2 l_0) \biggl(H_{0,1,0,1}(y) + 2 H_{0}(y) H_{0,1,1}(y) + 2 H_{1}(y) H_{0,0,1}(y)\biggr)
- (l_1 l - 3 l_2 l_0) H_{0,1}^2(y)
- 4 l_1 l H_{0}(y) H_{1}(y) H_{0,1}(y)
\nonumber\\
&&\quad{} + (l_1+l_2) (l_1+l_0) H_{0}^2(y) H_{1}^2(y)
- 2 l_2 l_0 \frac{\pi^2}{3} \bigl(H_{0,1}(y) - H_{0}(y) H_{1}(y)\bigr)
+ 4 \bigl(l_2 (l_1+l_2) - (l_1-l_2) l_0\bigr) \zeta_3 H_{1}(y)
\nonumber\\
&&\quad{} - \bigl(7 l_1 l + 4 (l_2^2 - l_2 l_0 + l_0^2)\bigr) \frac{\pi^4}{90}
\biggr] \varepsilon^4 + \mathcal{O}(\varepsilon^5)\,,
\nonumber\\
&&y^{l\varepsilon} F_2(1,y) = 1
+ 2 l_2 l \biggl(2 H_{0,1}(y)- 2 H_{0}(y) H_{1}(y) - H_{0}^2(y) - \frac{\pi^2}{3}\biggr) \varepsilon^2
\nonumber\\
&&{} - 4 l_2 l \biggl[
(l_1+l_2) \biggl(2 H_{0,1,1}(y)- 2 H_{1}(y) H_{0,1}(y) + H_{0}(y) H_{1}^2(y)
+ \frac{\pi^2}{3} \bigl(H_{0}(y) + H_{1}(y)\bigr)
\biggr)
- 2 (l_1-l_0) H_{0,0,1}(y)
\nonumber\\
&&\quad{} - 2 (l_2+l_0) \bigl(H_{0}(y) H_{0,1}(y) + \zeta_3\bigr)
+ (l_2+l) H_{0}^2(y) \biggl(H_{1}(y) + \frac{1}{3} H_{0}(y)\biggr)
\biggr] \varepsilon^3
\nonumber\\
&&{} + 4 l_2 l \biggl[
2 (l_1+l_2)^2 \biggl(
2 H_{0,1,1,1}(y) - 2 H_{1}(y) H_{0,1,1}(y) + H_{1}^2(y) H_{0,1}(y)
- \frac{1}{3} H_{0}(y) H_{1}^3(y) - \frac{\pi^2}{6} H_{1}^2(y)
\biggr)
\nonumber\\
&&\quad{} + 2 (l_1 l - l_2 l_0) \bigl(H_{0,1,0,1}(y) + 2 H_{1}(y) H_{0,0,1}(y)\bigr)
+ 4 (l_1+l_0)^2 H_{0,0,0,1}(y)
\nonumber\\
&&\quad{} - 4 \bigl(l_2 (l_1+l_2) - (l_1-l_2) l_0\bigr)
\bigl(H_{0}(y) H_{0,1,1}(y) - \zeta_3 \bigl(H_{0}(y) + H_{1}(y)\bigr)\bigr)
+ 4 (l_1 l_2 - l l_0) H_{0}(y) H_{0,0,1}(y)
\nonumber\\
&&\quad{} - \bigl(l_1 (l_1 + l_2 + 3 l_0) - l_2 l_0\bigr) H_{0,1}^2(y)
+ 2 \bigl((l_2+l_0)^2 + l_1 l_0\bigr) H_{0}^2(y) H_{0,1}(y)
+ 4 l_2 l H_{0}(y) H_{1}(y) H_{0,1}(y)
- (l_1+l_2) (l_2+l) H_{0}^2(y) H_{1}^2(y)
\nonumber\\
&&\quad{} - \frac{1}{6} \bigl((l_1+l_0)^2 + 3 l_2 l\bigr) H_{0}^3(y) \bigl(4 H_{1}(y) + H_{0}(y)\bigr)
+ \frac{\pi^2}{3} \bigl(
2 l_1 l_0 H_{0,1}(y)
- \bigl((l_1+l_2)^2 + l_1 l_0\bigr) H_{0}(y) \bigl(2 H_{1}(y) + H_{0}(y)\bigr)
\bigr)
\nonumber\\
&&\quad{} - \bigl(7 l_1 (l_1+l_2) + 4 l_2^2 + (12 l_1 + l_2) l_0 + 4 l_0^2\bigr) \frac{\pi^4}{90}
\biggr] \varepsilon^4 + \mathcal{O}(\varepsilon^5)\,.
\label{Fy}
\end{eqnarray}
\end{widetext}
This expansion can be straightforwardly extended to any order in $\varepsilon$.
We have also expanded the exact hypergeometric representations of $F_{1,2}(1,y)$
which follow from~(\ref{BBG}) up to $\varepsilon^3$ using HypExp~\cite{Huber:2005yg,Huber:2007dx}.
The results can be expressed via ordinary polylogarithms up to $\Li3$,
and agree with~(\ref{Fy}).

Any finite number of terms in the expansion of $F(1,y)$ in $\bar{y}=1-y$
can be straightforwardly obtained from~(\ref{BBG}):
\begin{eqnarray}
&&F_1(1,y) = 1 - \frac{\varepsilon l \bar{y}}{1+2(l_1+l_2)\varepsilon}
\left[1 - 2 (l_1-l_2) \varepsilon + \mathcal{O}(\bar{y})\right]\,,
\nonumber\\
&&F_2(1,y) = 1 + \frac{\varepsilon l \bar{y}}{1+2(l_1+l_2)\varepsilon}
\left[1 + 2 (l_1-l_2) \varepsilon + \mathcal{O}(\bar{y})\right]
\label{Ser1y}
\end{eqnarray}
(we have obtained them up to $\bar{y}^{20}$).
This expansion satisfies the differential equations~(\ref{dey1})
with the initial conditions~(\ref{Init}).
Expanding each coefficient of~(\ref{Ser1y}) in $\varepsilon$,
and each coefficient of~(\ref{Fy}) in $\bar{y}$,
we obtain two identical double expansions up to $\varepsilon^4$ and $\bar{y}^{20}$;
this is a strong check of our result~(\ref{Fy}).

\begin{figure}
\includegraphics{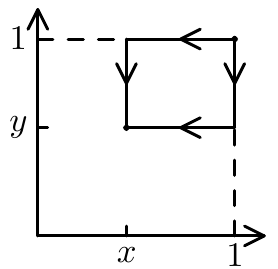}
\caption{Paths from $(1,1)$ to $(x,y)$.}
\label{paths}
\end{figure}

Finally, we discuss the general case.
Due to the symmetry relations~(\ref{Sym2})
it is sufficient to consider the region $x\le1$, $y\le1$.
We can solve the differential equations~(\ref{de2})
along one of the two paths in Fig.~\ref{paths}.
The result is a combination of products
of Goncharov polylogarithms~\cite{Goncharov:1998kja}
\[
G_{\underbrace{0,\ldots,0}_n}(x) = \frac{1}{n!} \log^n x\,,\quad
G_{a,\ldots}(x) = \int_0^x \frac{dt}{t-a} G_{\ldots}(t)
\]
of $\bar{x}=1-x$ and $\bar{y}=1-y$.
Numerical evaluation of Goncharov polylogarithms is available~\cite{Vollinga:2004sn}
in GiNaC~\cite{Bauer:2000cp}.
We make no efforts to express some of them
via harmonic polylogarithms of $x$ and $y$
because some Goncharov polylogarithms are bound to remain.
Using Libra we obtain
\begin{widetext}
\begin{eqnarray}
&&y^{-l\varepsilon} F_1(x,y) = 1
+ 2 l_1 \bigl\{l \bigl[G_{\bar{y},1}(\bar{x}) - G_{\hat{y},1}(\bar{x}) - 2 G_{0,1}(\bar{y})
+ G_{1}(\bar{y}) \bigl(G_{1}(\bar{x}) - G_{\bar{y}}(\bar{x}) - G_{\hat{y}}(\bar{x})\bigr)\bigr]
+ (l_1-l_2+l_0) G_{1,1}(\bar{x})\bigr\} \varepsilon^2
\nonumber\\
&&{} - 2 l_1 \bigl\{l \bigl[l_0 \bigl(2 (G_{\hat{y},0,1}(\bar{x}) - G_{\bar{y},0,1}(\bar{x}) + G_{\hat{y},2,1}(\bar{x}) - G_{\bar{y},2,1}(\bar{x}))
+ G_{\bar{y},\hat{y},1}(\bar{x}) - G_{\hat{y},\bar{y},1}(\bar{x})
+ G_{1}(\bar{y}) (G_{\bar{y},\hat{y}}(\bar{x}) + G_{\hat{y},\bar{y}}(\bar{x}) - G_{\hat{y},1}(\bar{x}))\bigr)
\nonumber\\
&&\qquad{} + (2l-l_0) \bigl(G_{\bar{y},\bar{y},1}(\bar{x}) - G_{\hat{y},\hat{y},1}(\bar{x})
- G_{1}(\bar{y}) (G_{\bar{y},\bar{y}}(\bar{x}) + G_{\hat{y},\hat{y}}(\bar{x}) - G_{\bar{y},1}(\bar{x}))\bigr)
\nonumber\\
&&\qquad{} + 2 (l_1+l_2) \bigl(G_{0,1}(\bar{y}) (G_{1}(\bar{x}) - G_{\bar{y}}(\bar{x}) - G_{\hat{y}}(\bar{x}))
- 2 G_{0,0,1}(\bar{y})\bigr)
+ (2l_1+l_0) \bigl(G_{1,\hat{y},1}(\bar{x}) + G_{1}(\bar{y}) G_{1,\hat{y}}(\bar{x})\bigr)
+ (2l_1-l_0) G_{\hat{y},1,1}(\bar{x})
\nonumber\\
&&\qquad{} - (2l_2-l_0) \bigl(G_{\bar{y},1,1}(\bar{x}) + G_{1,\bar{y},1}(\bar{x})
- G_{1}(\bar{y}) (G_{1,\bar{y}}(\bar{x}) - G_{1,1}(\bar{x}))\bigr)
+ 2 (l_1+l_0) \bigl(G_{1,1}(\bar{y}) (G_{1}(\bar{x}) - G_{\bar{y}}(\bar{x}) - G_{\hat{y}}(\bar{x}))
- 2 G_{0,1,1}(\bar{y})\bigr)\bigr]
\nonumber\\
&&\quad{} - 2 l_0 (l_1-l_2+l_0) \bigl(G_{1,0,1}(\bar{x}) + G_{1,2,1}(\bar{x})\bigr)
- \bigl(2(l_1^2-l_2^2)+(l_1+l_2)l_0-l_0^2\bigr) G_{1,1,1}(\bar{x})\bigr\} \varepsilon^3
+ \mathcal{O}(\varepsilon^4)
\nonumber\\
&&{} = 1
+ 2 l_1 \bigl\{l \bigl[G_{1}(\bar{x}) \bigl(G_{\bar{x}}(\bar{y}) - G_{\hat{x}}(\bar{y})\bigr)
- G_{\bar{x},1}(\bar{y})-G_{\hat{x},1}(\bar{y})\bigr]
+ (l_1-l_2+l_0) G_{1,1}(\bar{x})\bigr\} \varepsilon^2
\nonumber\\
&&{} - 2 l_1 \bigl\{l \bigl[l_0 \bigl(2 G_{2,1}(\bar{x}) (G_{\hat{x}}(\bar{y}) - G_{\bar{x}}(\bar{y}))
+ G_{1}(\bar{x}) (G_{\bar{x},\hat{x}}(\bar{y}) - G_{\hat{x},\bar{x}}(\bar{y}))
+ G_{\bar{x},\hat{x},1}(\bar{y}) + G_{\hat{x},\bar{x},1}(\bar{y})\bigr)
\nonumber\\
&&\qquad{} + (2l-l_0) \bigl(G_{1}(\bar{x}) (G_{\bar{x},\bar{x}}(\bar{y}) - G_{\hat{x},\hat{x}}(\bar{y}))
- G_{\bar{x},\bar{x},1}(\bar{y}) - G_{\hat{x},\hat{x},1}(\bar{y})\bigr)
+ 2 (l_1+l_2) G_{0,1}(\bar{x}) \bigl(G_{\bar{x}}(\bar{y}) - G_{\hat{x}}(\bar{y})\bigr)
+ 2 (l_1-l_2) G_{0,1,1}(\bar{x})
\nonumber\\
&&\qquad{} + (2l_1-l_0) G_{1,1}(\bar{x}) G_{\hat{x}}(\bar{y})
- (2l_2-l_0) G_{1,1}(\bar{x}) G_{\bar{x}}(\bar{y})
- 2 (l_1+l_0) \bigl(G_{\bar{x},1,1}(\bar{y}) + G_{\hat{x},1,1}(\bar{y})\bigr)\bigr]
\nonumber\\
&&\quad{} - 2 (l_1-l_2+l_0) \bigl[l_0 G_{1,2,1}(\bar{x}) - (l_1+l_2) G_{1,0,1}(\bar{x})\bigr]
- \bigl(2(l_1^2-l_2^2)+(l_1+l_2)l_0-l_0^2\bigr) G_{1,1,1}(\bar{x})\bigr\} \varepsilon^3
+ \mathcal{O}(\varepsilon^4)\,,
\nonumber\\
&&y^{l\varepsilon} F_2(x,y) = 1
+ 2 l_2 \bigl\{l \bigl[2 \bigl(G_{0,1}(\bar{y}) - G_{1,1}(\bar{y})\bigr)
- G_{\bar{y},1}(\bar{x}) + G_{\hat{y},1}(\bar{x})
- G_{1}(\bar{y}) \bigl(G_{1}(\bar{x}) - G_{\hat{y}}(\bar{x}) - G_{\bar{y}}(\bar{x})\bigr)\bigr]
- (l_1-l_2-l_0) G_{1,1}(\bar{x})\bigr\} \varepsilon^2
\nonumber\\
&&{} - 2 l_2 \bigl\{l \bigl[l_0 \bigl(G_{1}(\bar{y}) (G_{\bar{y},1}(\bar{x}) - G_{\bar{y},\hat{y}}(\bar{x}) - G_{\hat{y},\bar{y}}(\bar{x}))
+ 2 (G_{\bar{y},0,1}(\bar{x}) - G_{\hat{y},0,1}(\bar{x}) + G_{\bar{y},2,1}(\bar{x}) - G_{\hat{y},2,1}(\bar{x}))
- G_{\bar{y},\hat{y},1}(\bar{x}) + G_{\hat{y},\bar{y},1}(\bar{x})\bigr)
\nonumber\\
&&\qquad{} + (2l-l_0) \bigl(G_{1}(\bar{y}) (G_{\bar{y},\bar{y}}(\bar{x}) + G_{\hat{y},\hat{y}}(\bar{x}) - G_{\hat{y},1}(\bar{x}))
- G_{\bar{y},\bar{y},1}(\bar{x}) + G_{\hat{y},\hat{y},1}(\bar{x})\bigr)
\nonumber\\
&&\qquad{} + (l_1+l_2) \bigl(4 (G_{0,0,1}(\bar{y}) - G_{1,0,1}(\bar{y}))
- 2 G_{0,1}(\bar{y}) (G_{1}(\bar{x}) - G_{\bar{y}}(\bar{x}) - G_{\hat{y}}(\bar{x}))\bigr)
\nonumber\\
&&\qquad{} - (2l_1-l_0) \bigl(G_{1}(\bar{y}) (G_{1,\hat{y}}(\bar{x}) - G_{1,1}(\bar{x}))
+ G_{\hat{y},1,1}(\bar{x}) + G_{1,\hat{y},1}(\bar{x})\bigr)
- (2l_2+l_0) \bigl(G_{1}(\bar{y}) G_{1,\bar{y}}(\bar{x}) - G_{1,\bar{y},1}(\bar{x})\bigr)
\nonumber\\
&&\qquad{} + (2l_2-l_0) G_{\bar{y},1,1}(\bar{x})
+ 2 (l_2+l) \bigl(2 (G_{1,1,1}(\bar{y}) - G_{0,1,1}(\bar{y}))
+ G_{1,1}(\bar{y}) (G_{1}(\bar{x}) - G_{\bar{y}}(\bar{x}) - G_{\hat{y}}(\bar{x}))\bigr)\bigr]
\nonumber\\
&&\quad{} + 2 l_0 (l_1-l_2-l_0) \bigl(G_{1,0,1}(\bar{x}) + G_{1,2,1}(\bar{x})\bigr)
+ \bigl(2(l_1^2-l_2^2)-(l_1+l_2)l_0+l_0^2\bigr) G_{1,1,1}(\bar{x})\bigr\} \varepsilon^3
+ \mathcal{O}(\varepsilon^4)
\nonumber\\
&&{} = 1
+ 2 l_2 \bigl\{l \bigl[G_{1}(\bar{x}) \bigl(G_{\hat{x}}(\bar{y}) - G_{\bar{x}}(\bar{y})\bigr)
- 2 G_{1,1}(\bar{y}) + G_{\bar{x},1}(\bar{y}) + G_{\hat{x},1}(\bar{y})\bigr]
- (l_1-l_2-l_0) G_{1,1}(\bar{x})\bigr\} \varepsilon^2
\nonumber\\
&&{} - 2 l_2 \bigl\{2 l^2 G_{1}(\bar{x}) \bigl(G_{\bar{x},1}(\bar{y}) - G_{\hat{x},1}(\bar{y})\bigr)
+ l \bigl[l_0 \bigl(2 G_{2,1}(\bar{x}) (G_{\bar{x}}(\bar{y}) - G_{\hat{x}}(\bar{y}))
+ G_{1}(\bar{x}) (G_{\hat{x},\bar{x}}(\bar{y}) - G_{\bar{x},\hat{x}}(\bar{y}))
- G_{\bar{x},\hat{x},1}(\bar{y}) - G_{\hat{x},\bar{x},1}(\bar{y})\bigr)
\nonumber\\
&&\qquad{} + (2l-l_0) \bigl(G_{1}(\bar{x}) (G_{\hat{x},\hat{x}}(\bar{y}) - G_{\bar{x},\bar{x}}(\bar{y}))
+ G_{\bar{x},\bar{x},1}(\bar{y}) + G_{\hat{x},\hat{x},1}(\bar{y})\bigr)
\nonumber\\
&&\qquad{} + 2 (l_1+l_2) \bigl(G_{0,1}(\bar{x}) (G_{\hat{x}}(\bar{y}) - G_{\bar{x}}(\bar{y}))
+ G_{1}(\bar{x}) (G_{1,\bar{x}}(\bar{y}) - G_{1,\hat{x}}(\bar{y}))
- G_{1,\bar{x},1}(\bar{y}) - G_{1,\hat{x},1}(\bar{y})\bigr)
\nonumber\\
&&\qquad{} + 2 (l_1-l_2) \bigl(G_{1,1}(\bar{x}) G_{1}(\bar{y}) - G_{0,1,1}(\bar{x})\bigr)
- (2l_1-l_0) G_{1,1}(\bar{x}) G_{\hat{x}}(\bar{y})
+ (2l_2-l_0) G_{\bar{x}}(\bar{y}) G_{1,1}(\bar{x})
\nonumber\\
&&\qquad{} + 2 (l_2+l) \bigl(2 G_{1,1,1}(\bar{y}) - G_{\bar{x},1,1}(\bar{y}) - G_{\hat{x},1,1}(\bar{y})\bigr)\bigr]
\nonumber\\
&&\quad{} + 2 (l_1-l_2-l_0) \bigl[l_0 G_{1,2,1}(\bar{x}) - (l_1+l_2) G_{1,0,1}(\bar{x})\bigr]
+ \bigl(2(l_1^2-l_2^2)-(l_1+l_2)l_0+l_0^2\bigr) G_{1,1,1}(\bar{x})\bigr\} \varepsilon^3
+ \mathcal{O}(\varepsilon^4)\,,
\nonumber\\
&&F_3(x,y) = 2 G_{1}(\bar{x}) \varepsilon
+ 2 \bigl\{l \bigl[G_{1}(\bar{y}) \bigl(G_{\bar{y}}(\bar{x}) - G_{\hat{y}}(\bar{x})\bigr)
- G_{\bar{y},1}(\bar{x}) - G_{\hat{y},1}(\bar{x})\bigr]
+ 2 l_0 \bigl(G_{0,1}(\bar{x}) + G_{2,1}(\bar{x})\bigr)
+ (l_1+l_2-l_0) G_{1,1}(\bar{x})\bigr\} \varepsilon^2
\nonumber\\
&&{} - 2 \bigl\{l \bigl[l_0 \bigl(G_{1}(\bar{y}) (2 (G_{0,\hat{y}}(\bar{x}) - G_{0,\bar{y}}(\bar{x}) + G_{2,\hat{y}}(\bar{x}) - G_{2,\bar{y}}(\bar{x}))
- G_{\bar{y},\hat{y}}(\bar{x}) + G_{\hat{y},\bar{y}}(\bar{x}))
\nonumber\\
&&\qquad\quad{} + 2 (G_{0,\bar{y},1}(\bar{x}) + G_{0,\hat{y},1}(\bar{x}) + G_{2,\bar{y},1}(\bar{x}) + G_{2,\hat{y},1}(\bar{x}) + G_{\bar{y},0,1}(\bar{x}) + G_{\hat{y},0,1}(\bar{x}) + G_{\bar{y},2,1}(\bar{x}) + G_{\hat{y},2,1}(\bar{x}))
- G_{\bar{y},\hat{y},1}(\bar{x}) - G_{\hat{y},\bar{y},1}(\bar{x})\bigr)
\nonumber\\
&&\qquad{} + (2l-l_0) \bigl(G_{1}(\bar{y}) (G_{\bar{y},\bar{y}}(\bar{x}) - G_{\hat{y},\hat{y}}(\bar{x}))
- G_{\bar{y},\bar{y},1}(\bar{x}) - G_{\hat{y},\hat{y},1}(\bar{x})\bigr)
\nonumber\\
&&\qquad{} + (l_1+l_2) \bigl(2 G_{0,1}(\bar{y}) (G_{\bar{y}}(\bar{x}) - G_{\hat{y}}(\bar{x}))
- G_{1}(\bar{y}) (G_{\bar{y},1}(\bar{x}) - G_{\hat{y},1}(\bar{x}))\bigr)
+ (l_1-l_2) \bigl(2 G_{0,1}(\bar{y}) G_{1}(\bar{x}) - G_{1}(\bar{y}) G_{1,1}(\bar{x})\bigr)
\nonumber\\
&&\qquad{} + (2l_1-l_0) \bigl(G_{1}(\bar{y}) G_{1,\hat{y}}(\bar{x}) + G_{\hat{y},1,1}(\bar{x}) + G_{1,\hat{y},1}(\bar{x})\bigr)
- (2l_2-l_0) \bigl(G_{1}(\bar{y}) G_{1,\bar{y}}(\bar{x}) - G_{1,\bar{y},1}(\bar{x}) - G_{\bar{y},1,1}(\bar{x})\bigr)
\nonumber\\
&&\qquad{} - (l_1-l_2+l_0) G_{1}(\bar{x}) G_{1,1}(\bar{y})
+ 2 l_2 G_{1,1}(\bar{y}) \bigl(G_{\hat{y}}(\bar{x}) - G_{\bar{y}}(\bar{x})\bigr)\bigr]
\nonumber\\
&&\quad{} - 4 l_0^2 \bigl(G_{0,0,1}(\bar{x}) + G_{0,2,1}(\bar{x}) + G_{2,0,1}(\bar{x}) + G_{2,2,1}(\bar{x})\bigr)
- 2 l_0 (l_1+l_2-l_0) \bigl(G_{0,1,1}(\bar{x}) + G_{1,0,1}(\bar{x}) + G_{1,2,1}(\bar{x}) + G_{2,1,1}(\bar{x})\bigr)
\nonumber\\
&&\quad{} - \bigl(2(l_1^2+l_2^2)-(l_1+l_2)l_0+l_0^2\bigr) G_{1,1,1}(\bar{x})\bigr\} \varepsilon^3
+ \mathcal{O}(\varepsilon^4)
\nonumber\\
&&{} = 2 G_{1}(\bar{x}) \varepsilon
+ 2 \bigl\{l \bigl[G_{1}(\bar{x}) \bigl(G_{1}(\bar{y}) - G_{\bar{x}}(\bar{y}) - G_{\hat{x}}(\bar{y})\bigr)
+ G_{\bar{x},1}(\bar{y}) - G_{\hat{x},1}(\bar{y})\bigr]
+ (l_1+l_2-l_0) G_{1,1}(\bar{x}) - 2 (l_1+l_2) G_{0,1}(\bar{x})
+ 2 l_0 G_{2,1}(\bar{x})\bigr\} \varepsilon^2
\nonumber\\
&&{} - 2 \bigl\{l^2 \bigr[G_{1}(\bar{x}) \bigl(G_{1,\bar{x}}(\bar{y}) + G_{1,\hat{x}}(\bar{y}) + G_{\bar{x},1}(\bar{y})
+ G_{\hat{x},1}(\bar{y}) - G_{1,1}(\bar{y})\bigr)
- G_{1,\bar{x},1}(\bar{y}) + G_{1,\hat{x},1}(\bar{y})\bigr]
\nonumber\\
&&\quad{} + l \bigl[l_0 \bigl(2 G_{2,1}(\bar{x}) (G_{\bar{x}}(\bar{y}) + G_{\hat{x}}(\bar{y}) - G_{1}(\bar{y}))
- G_{1}(\bar{x}) (G_{\bar{x},\hat{x}}(\bar{y}) + G_{\hat{x},\bar{x}}(\bar{y}))
- G_{\bar{x},\hat{x},1}(\bar{y}) + G_{\hat{x},\bar{x},1}(\bar{y})\bigr)
\nonumber\\
&&\qquad{} - (2l-l_0) \bigl[G_{1}(\bar{x}) \bigl(G_{\bar{x},\bar{x}}(\bar{y}) + G_{\hat{x},\hat{x}}(\bar{y})\bigr)
- G_{\bar{x},\bar{x},1}(\bar{y}) + G_{\hat{x},\hat{x},1}(\bar{y})\bigr]
+ 2 (l_1+l_2) \bigl[G_{0,1}(\bar{x}) \bigl(G_{1}(\bar{y}) - G_{\hat{x}}(\bar{y})\bigr)
- G_{0,1}(\bar{x}) G_{\bar{x}}(\bar{y})\bigr]
\nonumber\\
&&\qquad{} + G_{1,1}(\bar{x}) \bigl[(2l_1-l_0) G_{\hat{x}}(\bar{y})
+ (2l_2-l_0) G_{\bar{x}}(\bar{y})
- (l_1+l_2-l_0) G_{1}(\bar{y})\bigr]
- 2 l_2 \bigl(G_{\bar{x},1,1}(\bar{y}) - G_{\hat{x},1,1}(\bar{y})\bigr)\bigr]
\nonumber\\
&&\quad{} - 4 (l_1+l_2)^2 G_{0,0,1}(\bar{x})
- 2 (l_1+l_2-l_0) \bigl[l_0 \bigl(G_{2,1,1}(\bar{x}) + G_{1,2,1}(\bar{x})\bigr)
- (l_1+l_2) \bigl(G_{1,0,1}(\bar{x}) + G_{0,1,1}(\bar{x})\bigr)\bigr]
\nonumber\\
&&\quad{} + 4 l_0 (l_1+l_2) (G_{2,0,1}(\bar{x}) + G_{0,2,1}(\bar{x}))
- 4 l_0^2 G_{2,2,1}(\bar{x})
- \bigl(2(l_1^2+l_2^2)-(l_1+l_2)l_0+l_0^2\bigr) G_{1,1,1}(\bar{x})\bigr\} \varepsilon^3
+ \mathcal{O}(\varepsilon^4)\,,
\label{Gen}
\end{eqnarray}
\end{widetext}
where $\hat{x}=1-x^{-1}$, $\hat{y}=1-y^{-1}$.
All Goncharov polylogarithms up to weight 2 can be expressed via $\Li2$ and logarithms.
This expansion can be straightforwardly extended to any order in $\varepsilon$.

I am grateful to R.\,N.~Lee for numerous consultations on Libra.
The work was supported by the Russian ministry of science and higher education.

\bibliography{hv}

\end{document}